\documentclass[twocolumn]{aastex62}

\newcommand{\Msolar}{{\rm M_{\odot}}}   % solar mass symbol
\newcommand{\MJup}{M_{\rm Jup}}

\received{January 1, 2018}
\revised{January 1, 2018}
\accepted{January 1, 2018}
\submitjournal{ApJ}

\shorttitle{The Spiral Morphology of Elias 2-27}
\shortauthors{Forgan et al.}

\begin{document}

%\title{\uppercase{Are Elias 2-27's spiral arms driven by self-gravity, or by a companion?\\ Lessons and warnings from comparative spiral morphology}}

\title{\uppercase{Are Elias 2-27's spiral arms driven by self-gravity, or by a companion?\\ a comparative spiral morphology study}}

\correspondingauthor{Duncan H. Forgan}
\email{dhf3@st-andrews.ac.uk}

\author[0000-0003-1175-4388]{Duncan H. Forgan}
\affiliation{SUPA, School of Physics and Astronomy, University of St Andrews, North Haugh, St Andrews KY16 9SS, UK}
\affiliation{Centre for Exoplanet Science, University of St Andrew, St Andrews, KY16 9SS, UK}

\author[0000-0003-1008-1142]{John D. Ilee}
\affil{Institute of Astronomy, University of Cambridge, Cambridge, CB3 0HA, USA}

\author[0000-0002-3984-9496]{Farzana Meru}
\affil{Institute of Astronomy, University of Cambridge, Cambridge, CB3 0HA, USA}
\affiliation{Department of Physics, University of Warwick, Gibbet Hill Road, Coventry, CV4 7AL, UK}
\affiliation{Centre for Exoplanets and Habitability, University of Warwick, Gibbet Hill Road, Coventry, CV4 7AL, UK}

%% Note that the \and command from previous versions of AASTeX is now
%% depreciated in this version as it is no longer necessary. AASTeX 
%% automatically takes care of all commas and "and"s between authors names.

%% AASTeX 6.2 has the new \collaboration and \nocollaboration commands to
%% provide the collaboration status of a group of authors. These commands 
%% can be used either before or after the list of corresponding authors. The
%% argument for \collaboration is the collaboration identifier. Authors are
%% encouraged to surround collaboration identifiers with ()s. The 
%% \nocollaboration command takes no argument and exists to indicate that
%% the nearby authors are not part of surrounding collaborations.

%% Mark off the abstract in the ``abstract'' environment. 
\begin{abstract}

The spiral waves detected in the protostellar disc surrounding Elias 2-27 have been suggested as evidence of the disc being gravitationally unstable.  However, previous work has shown that a massive, stable disc undergoing an encounter with a massive companion are also consistent with the observations.  We compare the spiral morphology of smoothed particle hydrodynamic simulations modelling both cases.  The gravitationally unstable disc produces symmetric, tightly wound spiral arms with constant pitch angle, as predicted by the literature. The companion disc's arms are asymmetric, with pitch angles that increase with radius.  However, these arms are not well-fitted by standard analytic expressions, due to the high disc mass and relatively low companion mass.  We note that differences (or indeed similarities) in morphology between pairs of spirals is a crucial discriminant between scenarios for Elias 2-27, and hence future studies \emph{must} fit spiral arms individually.  If Elias 2-27 continues to show symmetric tightly wound spiral arms in future observations, then we posit that it is the first observed example of a gravitationally unstable protostellar disc.

% Zhu et al functions don't fit!
% Note that asymmetry an important observable
% Make a prediction

\end{abstract}

%% Keywords should appear after the \end{abstract} command. 
%% See the online documentation for the full list of available subject
%% keywords and the rules for their use.

%% From the front matter, we move on to the body of the paper.
%% Sections are demarcated by \section and \subsection, respectively.
%% Observe the use of the LaTeX \label
%% command after the \subsection to give a symbolic KEY to the
%% subsection for cross-referencing in a \ref command.
%% You can use LaTeX's \ref and \label commands to keep track of
%% cross-references to sections, equations, tables, and figures.
%% That way, if you change the order of any elements, LaTeX will
%% automatically renumber them.
%%
%% We recommend that authors also use the natbib \citep
%% and \citet commands to identify citations.  The citations are
%% tied to the reference list via symbolic KEYs. The KEY corresponds
%% to the KEY in the \bibitem in the reference list below. 

\section{Introduction} 
\label{sec:intro}

%ALMA is amazing blah.

% Observations of self-gravitating protostellar discs are currently rare, as they are typically heavily embedded

%\noindent Self-gravitating protostellar discs are thought to be a typical consequence of star formation.  As a molecular cloud collapses to form a protostellar system, simulations commonly indicate that the local disc mass is equal to or in excess of the protostar mass at very early times (+++).  

%\noindent where $c_s$ is the sound speed of the disc gas, $\kappa_{\rm epi}$ is the epicyclic frequency (which is equal to the angular frequency $\Omega$ if the disc is Keplerian) and $\Sigma$ is the surface density.  Discs with $Q \sim 1$ are gravitationally unstable, particularly to non-axisymmetric perturbations.  These perturbations undergo swing amplification to produce spiral structures.

Spiral structures generated by gravitationally unstable protostellar discs play a crucial role in the evolution of protostars and the planetary systems they eventually host.  At the instant of a system's formation, the star mass and disc mass are comparable. This guarantees that the Toomre Parameter \citep{Toomre_stability1964}:

\begin{equation}
    Q = \frac{c_s \kappa_{\rm epi}}{\pi G \Sigma} \sim 1,
\end{equation}

\noindent where $c_s$ is the sound speed of the disc gas, $\kappa_{\rm epi}$ is the epicyclic frequency (which is equal to the angular frequency $\Omega$ if the disc is Keplerian) and $\Sigma$ is the surface density.  As such, discs which satisfy $Q \sim 1$ will be unstable to non-axisymmetric perturbations, which undergo swing amplification into spiral structures.

\smallskip

At early times, these structures are typically strong global modes, which can  achieve rapid accretion of the disc onto the star via non-local angular momentum transport \citep{Laughlin_Bodenheimer_t_rad,Lodato_Rice_massive_disc,Forgan_alpha}.  Under the appropriate conditions, spiral arms can fragment into gravitationally bound objects, representing a formation channel for low mass stars, brown dwarfs, gas giant planets and in some very rare cases, terrestrial planets \citep{Gammie_betacool,Rice_beta_condition,Stamatellos_BD_formation,Galvagni_pop_syn,Forgan_pop_syn,Forgan_popsyn2}.

\begin{figure*}
\centering
\includegraphics[width=0.49\textwidth]{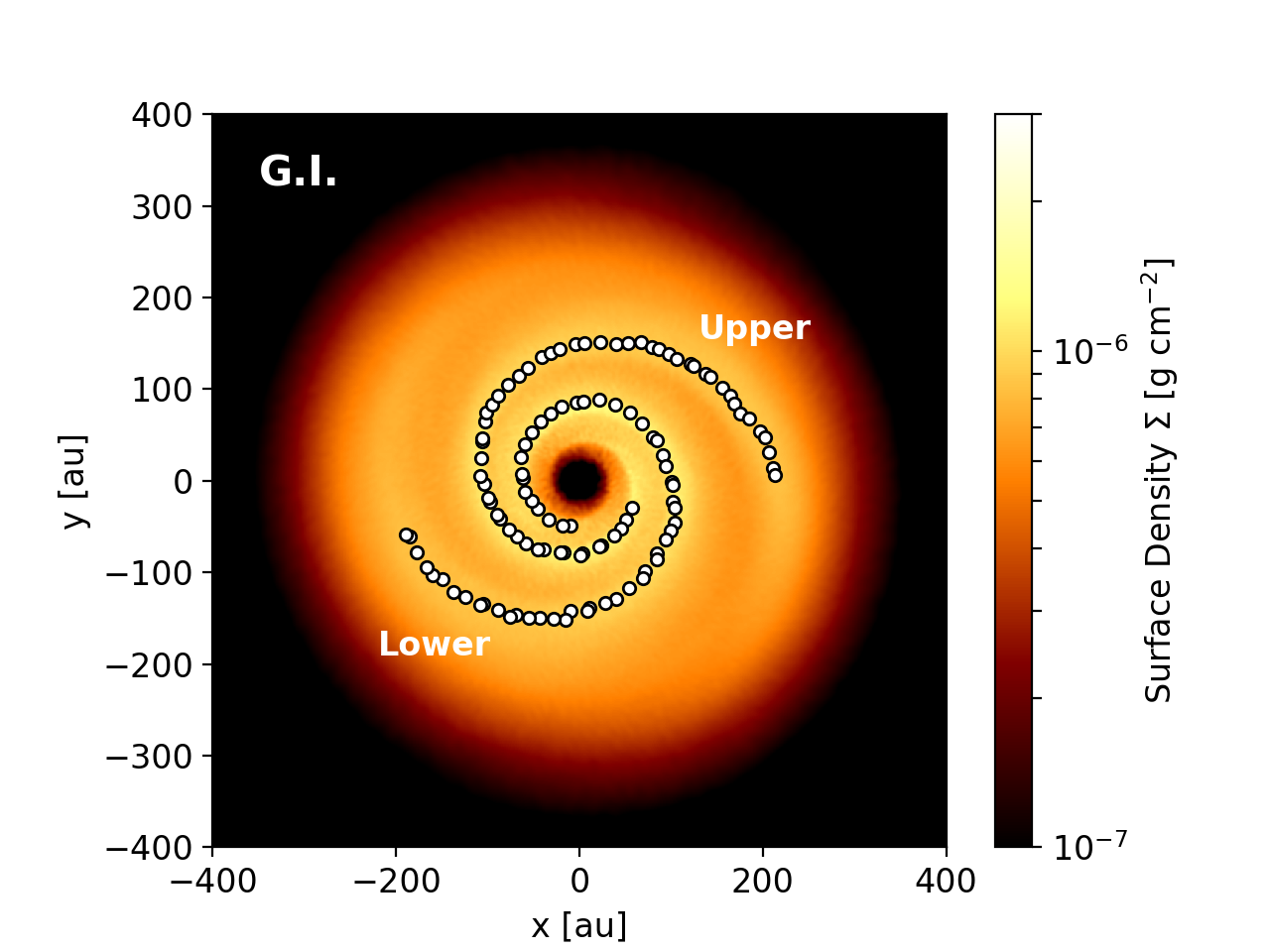}
\includegraphics[width=0.49\textwidth]{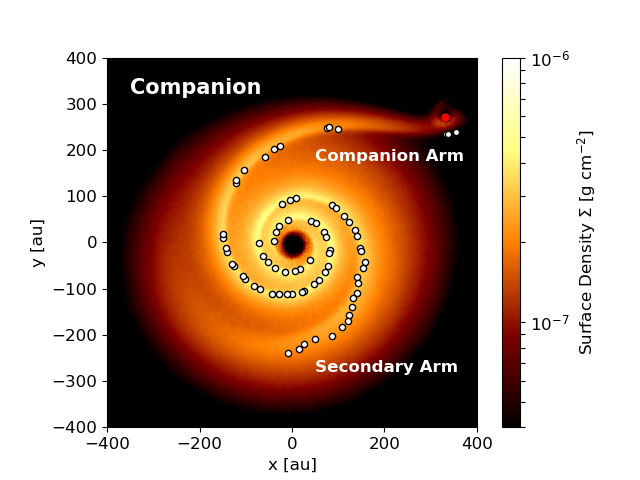}
\caption{Surface density of the gravitationally unstable disc (left) and the companion encounter disc (right) overlaid with the results of the {\sc tache} spiral identification algorithm (black crosses). Individual arms are identified as either 'upper' or 'lower' in each case.}
\label{fig:tache}
\end{figure*}

Constraining both disc fragmentation as a formation mechanism, and protostellar accretion in general, requires us to observe \emph{bona fide} gravitationally unstable protostellar discs in the wild.  For observational campaigns, there are (broadly) two approaches to determining whether a protostellar disc is gravitationally unstable - measuring its physical properties and determining $Q$, or studying its morphology for signs of spiral structure.  

\smallskip

Determining the physical properties of self-gravitating protostellar discs is a challenging endeavour, as the self-gravitating phase is brief due to rapid disc accretion.  As a result, self-gravitating discs remain heavily embedded inside their birth molecular cloud.  Further to this, their centrally condensed surface density profiles can frustrate efforts to measure disc masses through measuring continuum dust emission, due to large optical depths even at sub-millimetre wavelengths \citep{Greaves_Rice2010,Forgan_Rice_L1527IRS,Forgan_massivestars,evans_2017}.

\smallskip

The detection and characterisation of spiral structure, then, may yield a bias-free determination of whether a protostellar disc is gravitationally unstable.  Spirals in protostellar discs have recently come within the reach of observations.  Scattered light measurements have yielded several protostellar discs exhibiting spirals \citep[e.g.][]{Fukagawa_tidal_tail,Muto_disc_structure,Benisty2015,Dong2016}, but these observations only trace structure generated in the disc's upper layers.  Gravitationally unstable discs will drive structure in the bulk of the disc material, at all altitudes, down to the midplane. 

\smallskip

The arrival of the Atacama Large Millimetre Array (ALMA) has allowed exquisite spatially resolved observations of protostellar discs at sufficiently long wavelengths to probe spiral structure at the disc midplane.  A most striking recent example is Elias 2-27 -- a relatively young low mass star ($M_*\sim 0.5-0.6 \Msolar$, age $\sim$ 1 Myr; \citealt{rho_oph_spectraltype_age,rho_oph_accretion}), hosting a Class II circumstellar disc potentially exhibiting a large mass ($M_{\rm disc} \sim 0.04-0.14 \Msolar$, \citealt{Andrews_disc_properties,isella_2009_structure,Ricci_mm_cm_rhoOph}).  

\smallskip

\citet{Elias227_Science} presented ALMA observations that showed two large-scale, symmetric spiral arms.  In their study, both arms were simultaneously fitted to pure logarithmic spirals, with identical pitch angle $7.9^\circ \pm 0.4^\circ$.  Given both its relatively large disc-to-star mass ratio ($q\sim 0.06 - 0.3$) and the presence of spiral structure driven at the midplane, it has been suggested that Elias 2-27 is a gravitationally unstable disc system \citep{Tomida_Elias227_GI, meru_elias}.

\smallskip

Of course, gravitational instability (GI) is not the only mechanism that generates spiral structures.  Interactions between a disc and a companion generate tidally-driven arms, that (to the eye) can be very similar to arms driven in an isolated GI disc.  \citet{meru_elias} address this issue by running a suite of smoothed particle hydrodynamics (SPH) simulations of both isolated GI discs, and GI discs that are perturbed by a companion.  They show that both isolated and perturbed discs produce spiral structures that, when observed synthetically with ALMA, using the same unsharp masking technique as \citet{Elias227_Science}, produce images consistent with Elias 2-27's features.

\smallskip

\citet{Hall2018} also consider a range of synthetically observed isolated GI disc simulations, and show that if Elias 2-27 is an isolated GI disc, then its properties are tightly constrained, where slight changes to its physical properties result either in dissipation of the spirals or fragmentation.  This suggests that (\emph{a priori}) Elias 2-27 is less likely to be an isolated GI disc.

\smallskip

We therefore argue that comparative morphology studies are a crucial orthogonal tool to determine whether GI or a companion is driving spiral structure in a given protostellar disc.  In this work, we perform such a study on the simulations of \citet{meru_elias} to identify crucial differences in spiral morphologies between isolated and perturbed GI discs.  Our results offer several discriminants for determining the nature and origin of spiral structure in massive protostellar discs.

\section{METHODOLOGY}
\label{sec:numerics}

\subsection{Hydrodynamics}
\label{sec:hydro}

Our hydrodynamic simulations are fully described in \cite{meru_elias}, but for completeness we briefly reiterate some salient aspects.  The simulations are  performed using the three-dimensional Smoothed Particle Hydrodynamics code   ({\sc sphNG}) including heating due to work done and the radiative transfer of energy in the flux-limited diffusion limit \citep{WH_Bate_Monaghan2005,WH_Bate_science}.  A detailed description of the code can be found in \cite{Triggered} and \cite{meru_elias}.

\smallskip

Our reference model consists of a $0.5 \Msolar$ star, modelled as a sink particle, surrounded by a disc whose initial temperature ($T$) and surface density ($\Sigma$) vary with radius ($R$) as
\begin{eqnarray}
&T(R) = 13.4\,{\rm K} \left ( \frac{R}{200~{\rm au}} \right )^{-0.75}, \,\,{\rm and} \\
& \Sigma(R) = \Sigma_{\rm 0} \left ( \frac{R}{200 {\rm au}} \right )^{-0.75}
%\Sigma(R) = 5\,{\rm g\,cm^{-2}} \left ( \frac{R}{200~{\rm au}} \right )^{-0.75}
\label{eq:T_and_sigma}
\end{eqnarray}
respectively, between $R_{\rm in} = 10$~au and $R_{\rm out} = 350$~au.  The first simulation is a gravitationally unstable disc (GI), with $\Sigma_{\rm 0} = 6 \times 10^{-7} \Msolar \, \rm au^{-2}$, giving a total disc mass of 0.24 $\Msolar$, and an initial Toomre parameter $Q<1$ beyond $\sim$ 250 au.

In the second simulation (Companion), $\Sigma_{\rm 0} = 1.96 \times 10^{-7} \Msolar \,  \rm au^{-2}$, giving a total disc mass of $0.078 \Msolar$, and $Q>2$ at all radii.

We note that for the GI case the disc does evolve, such that $\Sigma \propto R^{-0.5}$.  For the Companion simulation we model an $8 \MJup$ companion initially located at 500~au that is allowed to freely interact with the disc, migrate and grow.  At the time when the simulation is analysed the companion is $\approx 10 \MJup$ and located at $\approx 425$ au from the central star.  The companion does not drive a gap, which is consistent with the gap opening criteria defined by both \citet{Lin_Papaloizou_TypeII} and \citet{Crida_gap}.  We note that the presence or absence of a gap makes little difference to our results.  Each disc is modelled using 250,000 SPH gas particles, and we assume that the gas and dust are well mixed \citep[see][for further details]{meru_elias}.

\smallskip

\begin{figure*}
\centering
\includegraphics[width=0.49\textwidth]{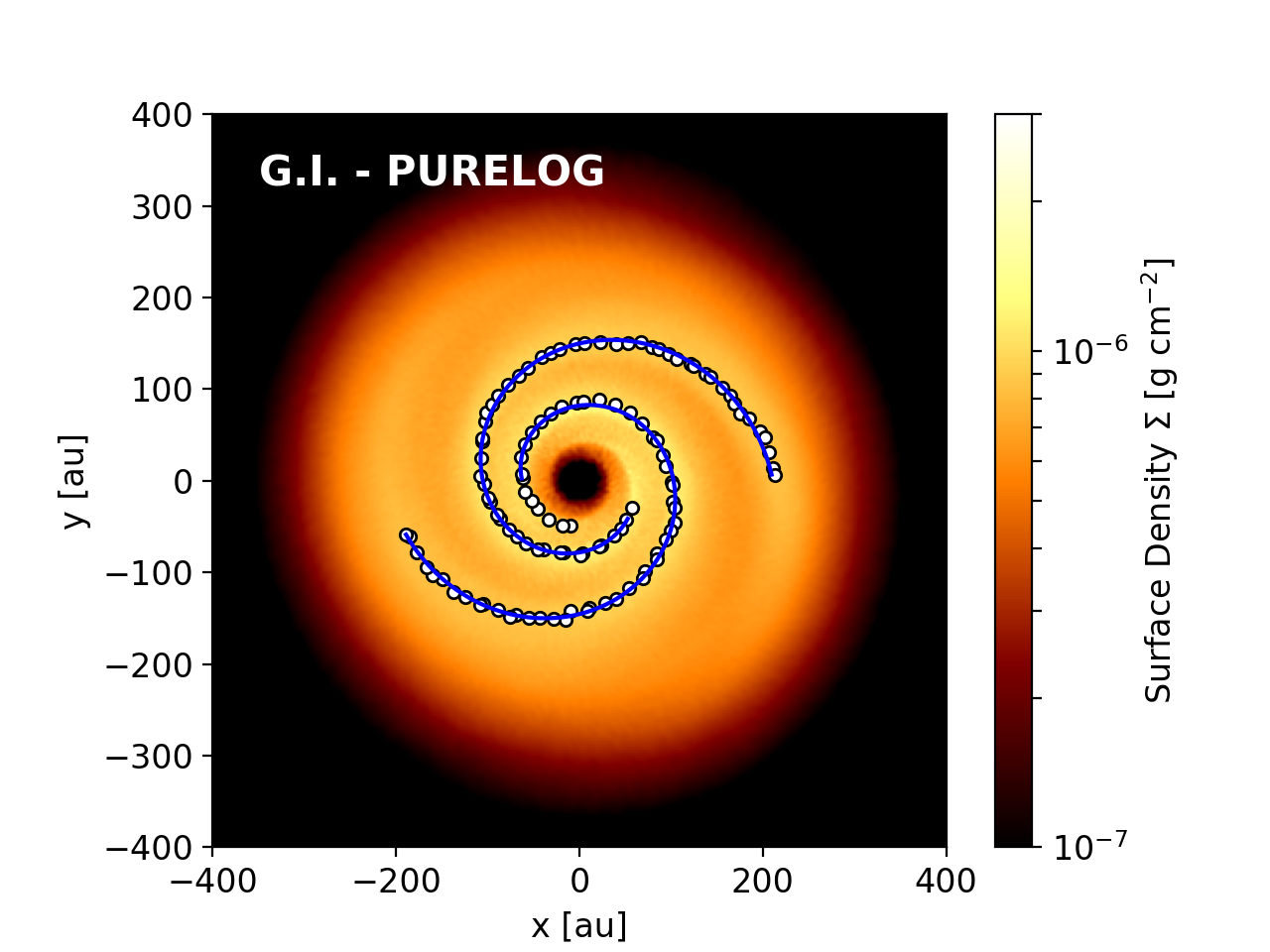}
\includegraphics[width=0.49\textwidth]{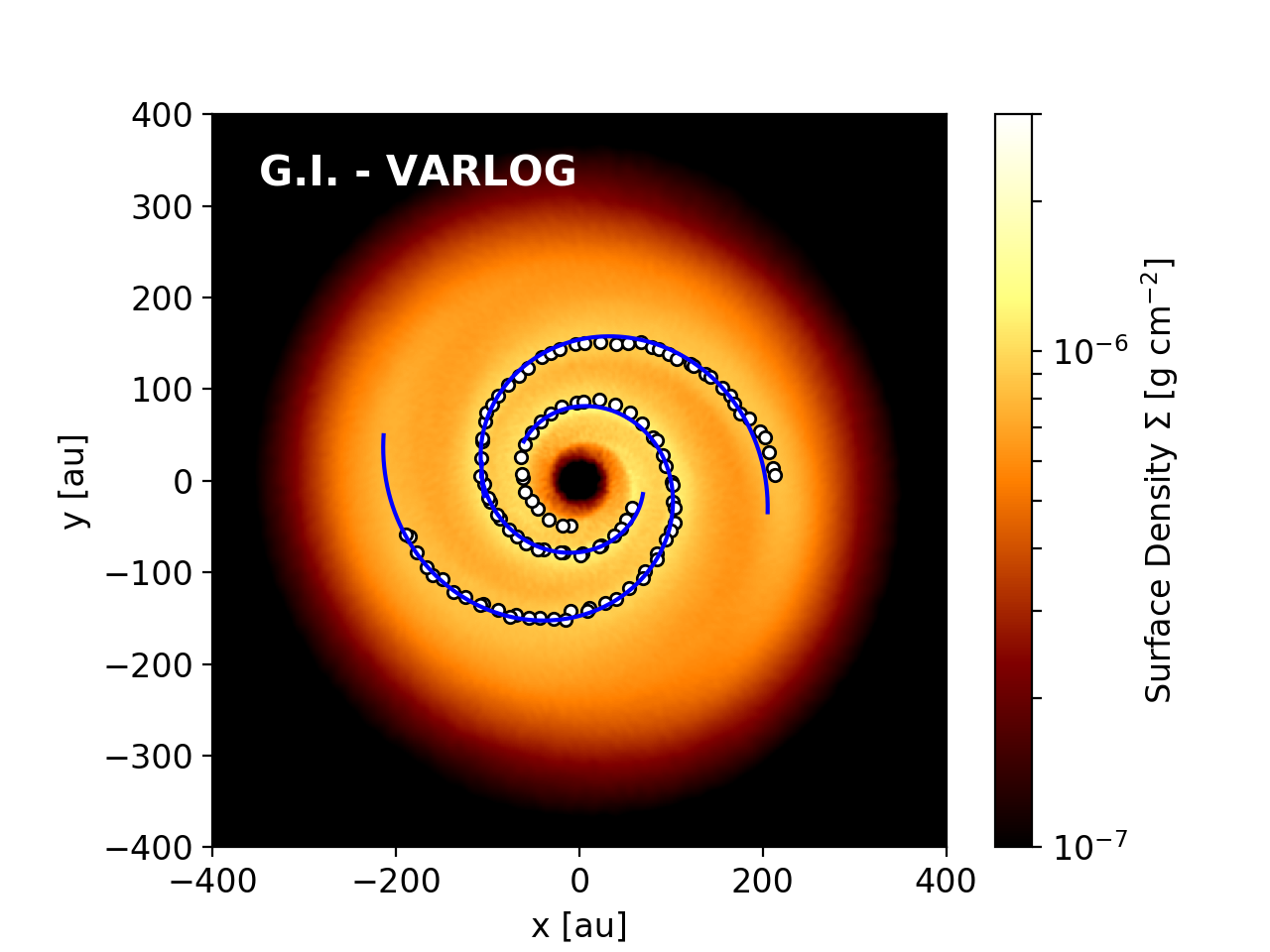}\\

\includegraphics[width=0.49\textwidth]{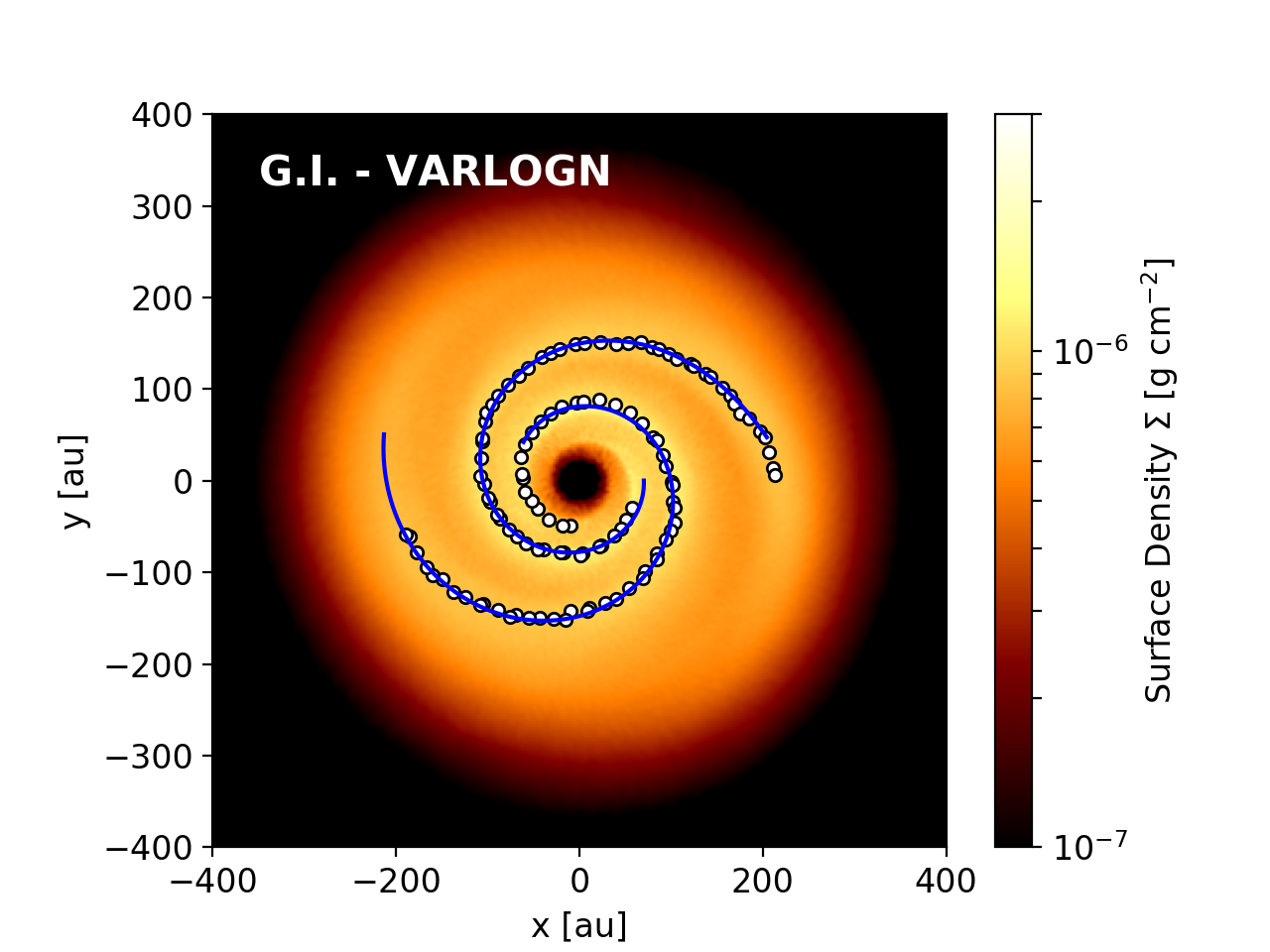}
\includegraphics[width=0.49\textwidth]{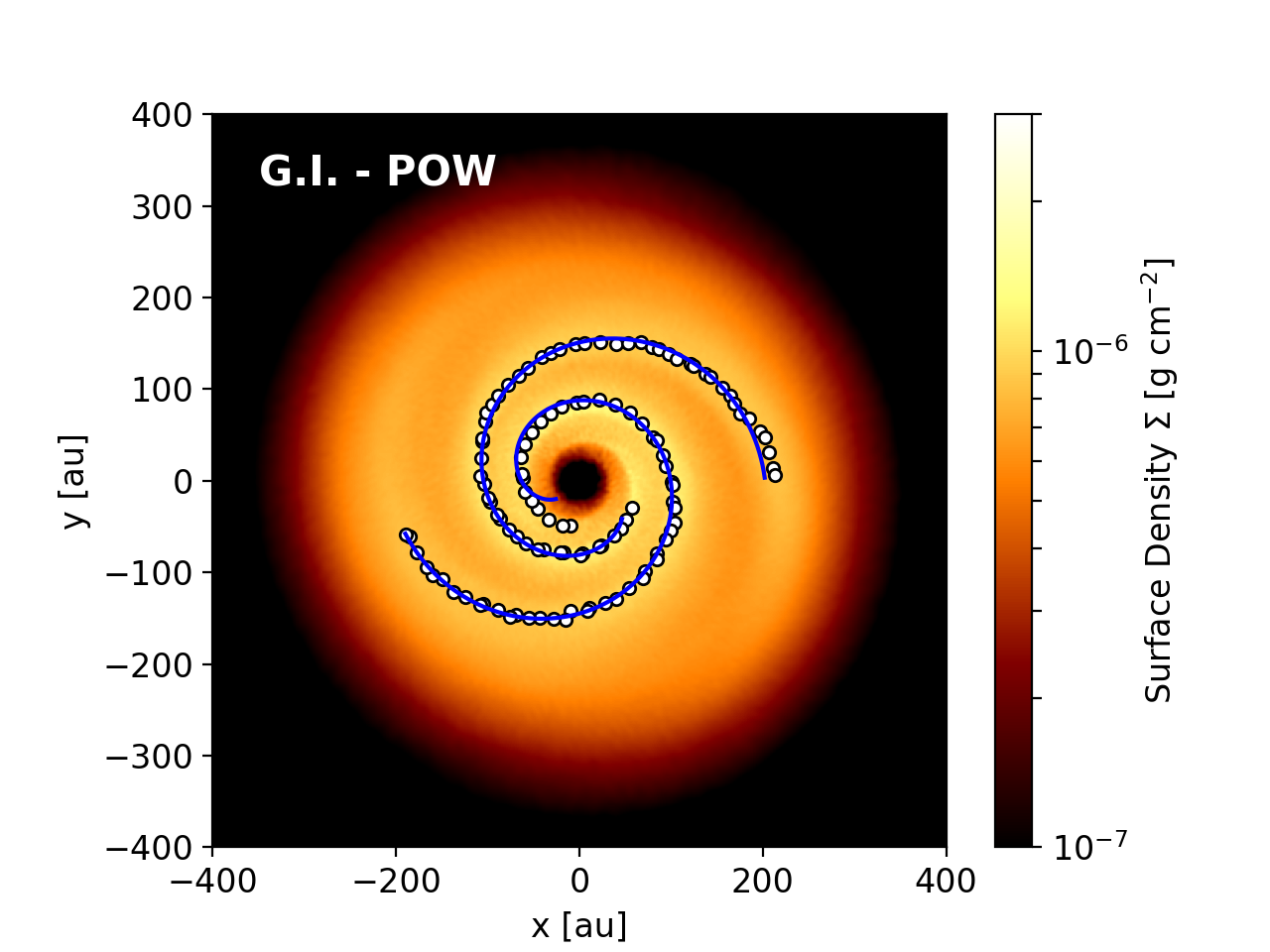}

\caption{The isolated GI disc simulation, overlaid with the {\sc tache} spine points and four different fitted spiral functions (see text).}
\label{fig:fits_GI}
\end{figure*}

\begin{figure*}
\centering
\includegraphics[width=0.49\textwidth]{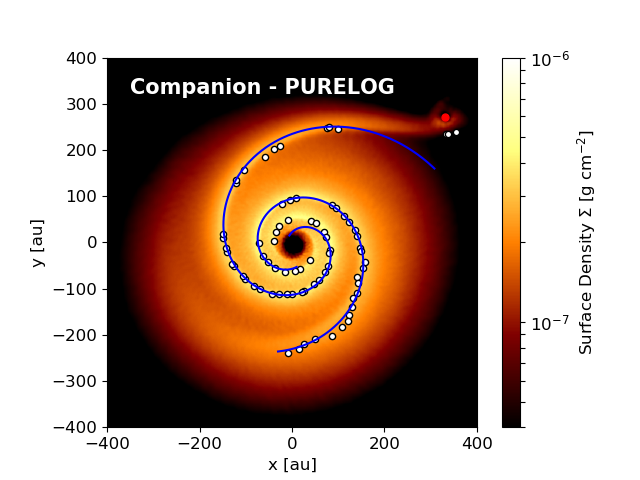}
\includegraphics[width=0.49\textwidth]{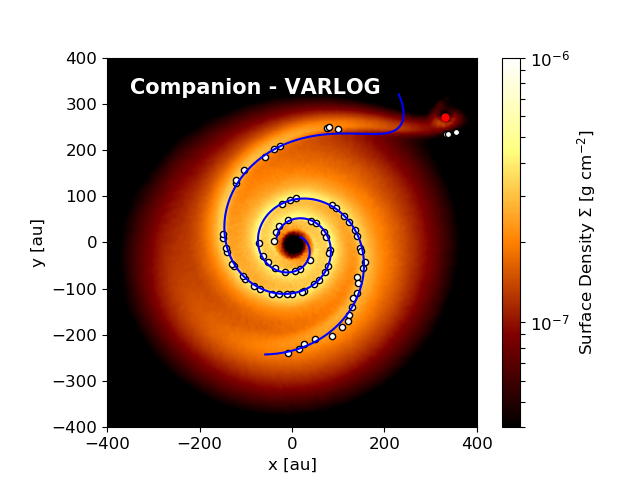} \\

\includegraphics[width=0.49\textwidth]{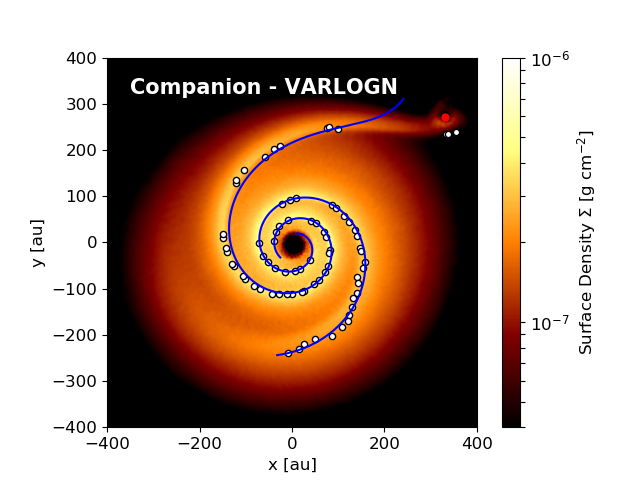}
\includegraphics[width=0.49\textwidth]{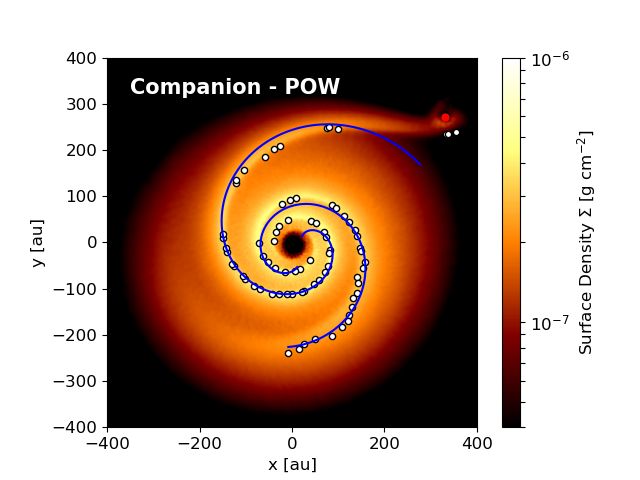}

\caption{As Figure \ref{fig:fits_GI}, but for the Companion simulation.  The location of the perturbing companion is shown in each panel by a red circle.}
\label{fig:fits_encounter}
\end{figure*}

\subsection{Spiral Detection \& Morphology \label{sec:spirals}}

We use the {\sc tache} code, which utilises tensor-classification of the simulations to determine which SPH particles reside in spiral structures \citep{Forgan_tensor2016,tache}.  Briefly, we compute the velocity shear tensor of each particle:

  \begin{equation}
\sigma_{ij} = -\frac{1}{2}\left(\frac{\partial v_i}{\partial x_j} + \frac{\partial v_j}{\partial x_i} \right).
\end{equation}

And then compute the tensor's eigenvalues.  The number of positive eigenvalues $E$ encodes information about the dimensionality of the flow.  For example, particles with $E=1$ indicate their motion is planar, consistent with motion in the undisturbed disc.  Particles with $E=2$ indicate 2D filamentary motions (in this case, spiral structure).  If the spiral structure is strong enough, particles near the centre of the arm will possess $E=3$ (3D collapse).

We can therefore identify particles belonging to the spiral using their $E$ value, and discard the other particles.  This allows us to trace the spine of the spiral structure (i.e. the location of maximum density), using a friends-of-friends-like algorithm, which yields a set of $(x,y)$ points for each individual spiral.  Each arm is then fitted separately via $\chi^2$ minimisation to a variety of spiral models (assuming a constant uncertainty of $\sigma=0.1$ AU for all points).  We use Nelder-Mead (amoeba) optimisation to obtain said minimum $\chi^2$, implemented via \texttt{scipy.optimize.minimize}.

\smallskip

Logarithmic spirals are commonly found in isolated GI discs and in discs driven by encounters with a companion:
\begin{equation}
    r = a e^{b\theta},
\end{equation}
where $r$ is cylindrical radius, and $\theta$ is the azimuthal angle. The $a$ parameter determines the initial distance of the spiral from the origin, and $b$ determines the winding properties of the arm.  The pitch angle of a logarithmic spiral 

\begin{equation}
    \phi = \left|r\frac{d\theta}{dr}\right|^{-1} = \arctan b.
\end{equation}

Pure logarithmic spirals (where $b=b_0$ is a constant) are typically found in simulations of isolated GI discs, with a constant $\phi \sim 10-15^\circ$ (for $q<0.5$ \citealt{Cossins_paper1,hall_2016,tache}).  We will label model fits of this type as PURELOG.

\smallskip

We also consider models where the pitch angle varies with radius, as is expected if the spiral is being driven by a companion \citep{goodman_2001,rafikov_2002,muto_2012}. For low mass companions ($M \lessapprox 1 \MJup$ in low mass discs ($M_d/M_*\lessapprox 0.1$), logarithmic spirals are typically found with the following function for $b$ \citep{zhu_2015},
\begin{equation}
b(r) = h_p\left(\frac{r_p}{r}\right)^{1+\eta}\frac{r^{\alpha}}{r^{\alpha} - r^{\alpha}_p},
\end{equation}
where $r_p$ is the orbital radius of the companion $\Omega \propto r^{-\alpha}$, $c_s \propto r^{-\eta}$, and $h_p$ is the aspect ratio of the disc at the location of the companion. 

\smallskip

We find that this function gives a very poor fit for both cases, as the spiral arms are deeply non-linear, due to the massive, self-gravitating nature of both discs (and the relatively high mass companion).  Despite the lack of theoretical guidance in this regime, we still consider the possibility that the pitch angle does vary with radius.  Instead, we fit a simpler function that describes a pitch angle that increases relatively slowly with radius (VARLOG),
\begin{equation}
    b(r) = b_0 + c\left(\frac{r-a}{a}\right)^{n}, \label{eq:bvarlog}
\end{equation}
where $b = b_0$ at $r=a$, and the PURELOG solution is recovered for $c=0$.  We also check specifically whether a linear dependence of $b$ with $r$ is sufficient by running fits with $n=1$ (VARLOGN).  Finally, to double check that the spirals do indeed possess a logarithmic form, we also fit a power spiral function (POW), where
\begin{equation}
    r = a \theta^n
\end{equation}
which also includes a radius-dependent pitch angle of the form

\begin{equation}
    \phi = n \left(\frac{r}{a}\right)^{-1/n},
\end{equation}
where for the special cases of $n=\pm 1$, the power spiral reduces to the Archimedean and hyperbolic spirals respectively (modulo constants describing the spiral origin).

\section{Results \& Discussion}

%Plots to go in:

% Plot of $\phi(r)$ vs r for Companion spirals (and best fit constant phi)

\noindent Figure \ref{fig:tache} shows the results of {\sc tache}'s spiral spine identification for the GI disc (left) and the Companion disc (right).  Note that for the Companion disc, the spine identification slightly misses the location of the companion.  This is due to the fact that the particles immediately surrounding the companion have a different tensor class to that of the spirals, and are thus removed from the spiral spine fitting.

In the following sections we attempt to fit these spiral spines using the spiral functions described previously.  A summary of our best fitting parameters for each case, along with the respective goodness-of-fit statistics, are given in Table \ref{tab:big}.

\begin{table*}
\centering
\caption{Best fitting parameters for both the G.I. and Companion simulation for the spiral functions described in Section \ref{sec:spirals}.  In each case, `up' and `low' refer to the upper and lower arms identified in Figure \ref{fig:tache}, respectively. \label{tab:big}}
\begin{tabular}{c|cccccccc|cccccccc|}
                    & \multicolumn{8}{c}{Gravitational Instability} & \multicolumn{8}{c}{Companion}     \\
                    
                    & \multicolumn{2}{c}{PURELOG}           &   \multicolumn{2}{c}{VARLOG}          &   \multicolumn{2}{c}{VARLOGN}             &   \multicolumn{2}{c}{POW}             &            
                      \multicolumn{2}{c}{PURELOG}             &   \multicolumn{2}{c}{VARLOG}          &   \multicolumn{2}{c}{VARLOGN}             &   \multicolumn{2}{c}{POW}             \\
                    
                    & Up     & Low & Up     & Low  & Up    & Low     & Up     & Low &
                      Up     & Low & Up     & Low  & Up    & Low     & Up     & Low                 \\
\hline
    $a$             &  29.2   & 56.9  &  111.8  &  56.2  &  110.3  & 56.8  & 55.3  & 11.5           %GI row
                    &  64.2   & 33.0  &  39.4    &  73.3 &   46.5   & 38.5 &  8.4  & 1.9 \\  %Encounter row
    $b_0$           &  0.208    & 0.205    & 0.191 & 0.244   &  0.123  &  0.216 & $-$       & $-$      %GI
                    &  0.288    & 0.243    & 0.208 & 0.225   &  0.148  &  0.201 & $-$       & $-$   \\ %encounters
    $\phi$ [$^\circ$]          &  11.75 &  11.59 & $-$       & $-$    & $-$      & $-$       & $-$       & $-$   
                    &  17.0 & 14.2  & $-$       & $-$    & $-$      & $-$       & $-$       & $-$   \\
    $c$             &  $-$      & $-$     & -0.009  &  -0.029     & 0.1      & -0.006   & $-$       & $-$   
                    &  $-$      & $-$     & 0.0003 & 0.0002 & 0.014 &  0.005    & $-$       & $-$   \\
    $n$             &  $-$      & $-$    & -0.49  & 0.35   & 1.0     & 1.0       &  0.67    & 1.26       
                    &  $-$      & $-$   &  2.40    & 1.92   & 1.0     & 1.0       &   1.64   &  1.14      \\
\hline
    $\chi^{2}$      &  1.98    & 3.98  & 3.63 & 3.22  & 3.02   & 3.26     & 3.96     & 4.13       
                    &  7.80    & 4.52  & 3.54 & 3.85  & 4.29   & 2.39     & 9.91     & 5.03       \\
 
\end{tabular}
\end{table*}

\subsection{Isolated GI disc \label{sec:GIdiscresults}}

\noindent We find that for the isolated GI disc, pure logarithmic spirals of constant pitch angle (PURELOG) deliver a very good fit to the data.  Other spiral functions yield reasonable fits to the arms (Figure \ref{fig:fits_GI}), but all yield poorer $\chi^2$ values than PURELOG for the upper arm, and deliver little change on the lower arm.  Power spiral fits (POW) produce quite poor fits at the inner and outer regions, strongly indicating that the arm's pitch angles are indeed constant with radius, as is the case for a PURELOG spiral.  VARLOG/VARLOGN fits require extra parameters and fail to yield better fits, again indicating that a constant pitch angle is the simplest and most effective model fit to these arms.  

\smallskip

Such a result is in accordance with our expectation that spiral arms in GI discs are density waves, and should hence be pure logarithmic spirals \citep{Cossins_paper1,hall_2016}.  It is also in accord with {\sc tache}'s previous applications to GI discs with disc-to-star mass ratios less than 0.5 \citep{tache}.  Note that the derived pitch angles for each arm are within 0.2$^\circ$ (1.4\%) of each other.  A hallmark of isolated GI discs are arms with extremely similar (if not identical) pitch angles.  

\smallskip

A naive conclusion might therefore be that Elias 2-27 is an isolated GI disc, as it too possesses symmetric arms with low pitch angle.  However, we must note that companion-driven simulations yield unsharp mask images that also appear to possess symmetric, tightly wound arms \citep{meru_elias}, and that Elias 2-27's arm symmetry may be an artifact of the fitting procedure (see section \ref{sec:implications}).  As we will see in the following section, the ``ground truth'' morphology of companion-driven spirals is markedly different.

\subsection{Companion disc}

\noindent We can see from Table \ref{tab:big} that PURELOG fits for the companion-driven arms are significantly poorer compared to the VARLOG/VARLOGN fits.  The minimum $\chi^2$ solution for PURELOG achieves a reasonable fit at intermediate radii, but at the cost of poorly fitting points in the inner and outer disc.  This is heavily indicative that the pitch angle does indeed vary with radius, as is expected for companion-driven arms \citep{goodman_2001,rafikov_2002,muto_2012,zhu_2015}.

\smallskip

Notably, while the companion-hosting arm is poorly fitted, the secondary arm is much better fitted, with both arms possessing larger pitch angles than the GI run.  The companion arm has a pitch angle of almost $17^\circ$, which is (just) beyond the 10-15$^\circ$ range expected for isolated GI discs \citep{Cossins_paper1}.  The secondary arm, at $\phi\approx 15^\circ$, therefore appears similar to a typical GI density wave, making it less easy to distinguish.

\smallskip

It is worth noting that analysing subsequent timesteps of the companion simulation yields a range of derived pitch angles for PURELOG, in some cases as large as $\sim 30^\circ$ for the companion arm.  We find throughout that the difference in pitch angles between companion and secondary arm are much larger than determined  for pairs of arms in the GI disc (section \ref{sec:GIdiscresults}). 

\smallskip

We verify that the spiral form is indeed logarithmic by considering the POW fits.  We find that we cannot improve our fits (indeed, they are slightly worse).  Again, we find that the best fit power spiral fails to correctly track the location of the companion.

\smallskip

The best fit to the companion spirals are from the VARLOG/VARLOGN fits.  For both fixed $n$ and varying $n$, we find fits to the companion arm that have  $\chi^2$ two to three times smaller than the PURELOG and POW fits.  Allowing $b$ to vary constrains the inner spiral much more effectively than the PURELOG fit, and from this analysis it seems clear that the companion arms are indeed logarithmic spirals, not power spirals.  However, in contrast to the GI case, the logarithmic spirals are better fitted with a varying pitch angle as opposed to a constant pitch angle.  

\smallskip

It is notable that for the VARLOG/VARLOGN fits our functional form for $b$ results in a minimum in $\theta(r)$:
\begin{equation}
    \theta(r) = \frac{1}{b(r)} \ln \left(\frac{r}{a}\right),
\end{equation}
which occurs at

\begin{equation}
    \frac{d\theta}{dr} = \frac{1}{br} - \frac{\theta}{b}\frac{db}{dr} = 0.
\end{equation}
As 
\begin{equation}
\frac{db}{dr} = nc\left(\frac{r-a}{a}\right)^{n} \frac{1}{r-a} = n\frac{b-b_0}{r-a}
\end{equation}
We therefore find a minimum in $\theta(r)$ at
\begin{equation}
    \theta =  \left(r \frac{db}{dr}\right)^{-1} = \frac{r-a}{r}\frac{1}{n(b-b_0)}.
\end{equation}
Hence we see that beyond this minimum, the fitted spiral function turns away from the companion.  Note that attempts to fit this function on subsequent timesteps delivers a good fit to the companion arm, and not the secondary arm.  Essentially, the length of the arm determines whether VARLOG/VARLOGN fits will capture the entire arm, or turn away at  $\frac{d\theta}{dr} = 0$.

\smallskip

This is an important issue for attempts to fit spiral structure driven by companions.  Despite {\sc tache} being able to roughly locate the companion inside the spiral structure, attempts to fit the structure with typical spiral functions uniformly fail to locate the companion correctly.  We therefore urge caution when attempting to determine the location of an unseen companion using spiral structure alone.  If the companion mass is much less than $\sim$0.02 times the stellar mass, the two arms can be significantly asymmetric, and the companion location can be determined using the two spiral arms alone \citep{Fung2015}.

\subsection{Implications for Elias 2-27 \label{sec:implications}}

\noindent Both the GI and Companion simulations, when observed synthetically and subjected to the same unsharp mask imaging techniques as carried out by \citet{Elias227_Science}, broadly reproduce the observed spiral morphology of Elias 2-27 \citep{meru_elias}.  However, our results identify key  discriminators between the ``ground-truth'' spiral morphologies of the two cases.

If Elias 2-27 is an isolated, gravitationally unstable protostellar disc, our simulations predict symmetric, logarithmic spiral arms of constant pitch $\phi \sim 12^\circ$, consistent with expectations from density wave theory \citep{LinShu1964,BertinLin1996}, which are appropriate as for this case the disc-to-star mass ratio $q \lessapprox 0.5$ \citep{Cossins_paper1,Forgan_alpha,hall_2016}.  This is slightly larger than the measured pitch of $\sim 8^\circ$ from \citet{Elias227_Science}, but we should be encouraged by the fact that our simulations produce a similar pitch angle without a great deal of tuning (as the simulations by Meru et al were not intended to reproduce the exact ALMA image, but were testing which scenarios produced morphologies that were consistent with the observations)

If Elias 2-27 is stable against GI, but undergoing encounters with an external companion, our simulations show it should produce asymmetric arms which have pitch angles that vary with radius, and a larger mean pitch overall.  Due to the relatively large disc mass in both simulations (and the relatively massive companion in the Companion simulation), we find that the typical expressions for companion-driven arms \citep[e.g.][]{zhu_2015} are a poor fit for the spirals.  We also find that other spiral functions (such as the power spiral) are generally a worse fit to the data.

The current ALMA observations indicate that Elias 2-27 does possess tightly wound, symmetric logarithmic spiral structure, suggesting that it is in fact a GI disc.  However, we note that \citet{Elias227_Science}'s fits to the arms assume symmetry, precluding the study of an important observable for determining the origin of spiral structure.  We recommend that future observations of Elias 2-27 (and other discs with midplane-driven spiral arms) conduct fitting of individual arms rather than a single, simultaneous fit to all arms.

%%%%%%%%%%%%%%%%%%%%%%%%%%%%%%%%%%%%%%%%%%%%%%%%%%%%%%%%%%%%%%%%%%%%
\section{Conclusions}
\label{sec:conc}
%%%%%%%%%%%%%%%%%%%%%%%%%%%%%%%%%%%%%%%%%%%%%%%%%%%%%%%%%%%%%%%%%%%%

% TAKE HOME MESSAGES
% Even if fitting with a logarithmic spiral, best fit pitch angles are going to be much much wider for encounter runs due to their increasing phi(r)

% Arms can have different fits for companion runs.  GI runs will fit well to log spirals with similar phi

% If Elias 2-27 is symmetric with tightly wound spirals (<15 degrees), most likely a GI disc

\noindent In this paper, we have conducted a comparative spiral morphology study on massive protostellar disc simulations tuned to reproduce the observed spiral structure in the protostellar disc Elias 2-27 \citep{Elias227_Science}.  One simulation presented an isolated disc with $Q \lessapprox 1$ resulting in gravitational instability (GI), the other presented a disc with $Q > 2$, undergoing perturbations during an encounter with a 10 Jupiter mass companion (Companion).  

Using the {\sc tache} algorithm \citep{tache} on the simulation data, we identify the spine points associated with individual spiral arms for both simulations, and fit these spine points to a variety of spiral functions.  Our results show that the two simulations have markedly different ``ground truth'' spiral morphology, and yet produce similar ``observed'' morphologies under unsharp masking \citep{meru_elias}.  

From this study, and from previous work, we identify several key discriminators between GI and companion-driven spiral structure for the case of Elias 2-27.  The GI disc exhibits pure, symmetric logarithmic spirals of constant pitch angle, whereas the Companion disc shows asymmetric logarithmic spirals, with a pitch angle that varies with radius.

In particular, we show that asymmetry between spiral arms is a key observable, and as such observers \emph{must} attempt to fit spirals individually, rather than assuming a single set of fit parameters for all spirals.  It is worth noting that arm asymmetry is sensitive to the orbital phase of the companion, and this should be factored into any predictions based purely on asymmetry.

In conjunction with \citet{meru_elias}, this study demonstrates that the current ALMA observations do not yet differentiate between the spiral structures observed in the GI and Companion simulations for Elias 2-27.  This highlights the need for synthetic observations as a tool to both evaluate numerical simulations and to interpret real observations \citep[cf][]{Haworth2017}.

In summary: we recommend further high resolution observations of Elias 2-27 to determine the source of its spiral structure. Between the two scenarios tested here, our simulations predict morphological differences on scales of 10-20au in the outer regions of the disc.  Therefore, future observations  of Elias 2-27 with angular resolutions of tens of milli-arcseconds have the potential to determine the  source  of its  spiral  structure.  We  predict  that  if  these  observations continue to show tightly wound, symmetric spiral structure,  then  Elias  2-27  is  indeed  the  first  observed self-gravitating protostellar disc.

\acknowledgments 

DHF gratefully acknowledges support from the ECOGAL project, grant agreement 291227, funded by the European Research Council under ERC-2011-ADG.  JDI and FM acknowledge support from the DISCSIM project, grant agreement 341137 under  ERC-2013-ADG.   FM also acknowledges   support  from   The  Leverhulme   Trust, the Isaac Newton Trust and the Royal Society Dorothy Hodgkin Fellowship.    The authors warmly thank the anonymous referee for their comments which helped to clarify the manuscript. This work used the Darwin DiRAC HPC cluster at the University of Cambridge, and the Cambridge COSMOS SMP system funded by ST/J005673/1, ST/H008586/1 and ST/K00333X/1 grants.  This research made use of NASA's Astrophysics Data System Bibliographic Services.

%% To help institutions obtain information on the effectiveness of their 
%% telescopes the AAS Journals has created a group of keywords for telescope 
%% facilities.
%
%% Following the acknowledgments section, use the following syntax and the
%% \facility{} or \facilities{} macros to list the keywords of facilities used 
%% in the research for the paper.  Each keyword is check against the master 
%% list during copy editing.  Individual instruments can be provided in 
%% parentheses, after the keyword, but they are not verified.

\vspace{5mm}
\facilities{Darwin DiRAC HPC cluster, University of Cambridge}

%% Similar to \facility{}, there is the optional \software command to allow 
%% authors a place to specify which programs were used during the creation of 
%% the manusscript. Authors should list each code and include either a
%% citation or url to the code inside ()s when available.

\software{\\{\sc tache}: \url{https://github.com/dh4gan/tache} \\{\sc scipy}: \url{http://www.scipy.org}}

\bibliographystyle{aasjournal}
\bibliography{allpapers}

\begin{thebibliography}{}
\expandafter\ifx\csname natexlab\endcsname\relax\def\natexlab#1{#1}\fi
\providecommand{\url}[1]{\href{#1}{#1}}

\bibitem[{{Andrews} {et~al.}(2009){Andrews}, {Wilner}, {Hughes}, {Qi}, \&
  {Dullemond}}]{Andrews_disc_properties}
{Andrews}, S.~M., {Wilner}, D.~J., {Hughes}, A.~M., {Qi}, C., \& {Dullemond},
  C.~P. 2009, \apj, 700, 1502

\bibitem[{{Benisty} {et~al.}(2015){Benisty}, {Juhasz}, {Boccaletti},
  {Avenhaus}, {Milli}, {Thalmann}, {Dominik}, {Pinilla}, {Buenzli}, {Pohl},
  {Beuzit}, {Birnstiel}, {de Boer}, {Bonnefoy}, {Chauvin}, {Christiaens},
  {Garufi}, {Grady}, {Henning}, {Huelamo}, {Isella}, {Langlois}, {M{\'e}nard},
  {Mouillet}, {Olofsson}, {Pantin}, {Pinte}, \& {Pueyo}}]{Benisty2015}
{Benisty}, M., {Juhasz}, A., {Boccaletti}, A., {et~al.} 2015, \aap, 578, L6

\bibitem[{{Bertin} \& {Lin}(1996)}]{BertinLin1996}
{Bertin}, G., \& {Lin}, C.~C. 1996, {Spiral structure in galaxies a density
  wave theory}, ISBN0262023962

\bibitem[{{Cossins} {et~al.}(2009){Cossins}, {Lodato}, \&
  {Clarke}}]{Cossins_paper1}
{Cossins}, P., {Lodato}, G., \& {Clarke}, C.~J. 2009, \mnras, 393, 1157

\bibitem[{{Crida} {et~al.}(2006){Crida}, {Morbidelli}, \& {Masset}}]{Crida_gap}
{Crida}, A., {Morbidelli}, A., \& {Masset}, F. 2006, \icarus, 181, 587

\bibitem[{Dong {et~al.}(2016)Dong, Zhu, Fung, Rafikov, Chiang, \&
  Wagner}]{Dong2016}
Dong, R., Zhu, Z., Fung, J., {et~al.} 2016, The Astrophysical Journal Letters,
  816, L12

\bibitem[{{Evans} {et~al.}(2017){Evans}, {Ilee}, {Hartquist}, {Caselli}, {Sz{\H
  u}cs}, {Purser}, {Boley}, {Durisen}, \& {Rawlings}}]{evans_2017}
{Evans}, M.~G., {Ilee}, J.~D., {Hartquist}, T.~W., {et~al.} 2017, \mnras, 470,
  1828

\bibitem[{{Forgan} {et~al.}(2016{\natexlab{a}}){Forgan}, {Bonnell}, {Lucas}, \&
  {Rice}}]{Forgan_tensor2016}
{Forgan}, D., {Bonnell}, I., {Lucas}, W., \& {Rice}, K. 2016{\natexlab{a}},
  \mnras, 457, 2501

\bibitem[{{Forgan} \& {Rice}(2013{\natexlab{a}})}]{Forgan_pop_syn}
{Forgan}, D., \& {Rice}, K. 2013{\natexlab{a}}, \mnras, 432, 3168

\bibitem[{{Forgan} \& {Rice}(2013{\natexlab{b}})}]{Forgan_Rice_L1527IRS}
---. 2013{\natexlab{b}}, \mnras, 433, 1796

\bibitem[{{Forgan} {et~al.}(2011){Forgan}, {Rice}, {Cossins}, \&
  {Lodato}}]{Forgan_alpha}
{Forgan}, D., {Rice}, K., {Cossins}, P., \& {Lodato}, G. 2011, \mnras, 410, 994

\bibitem[{{Forgan} {et~al.}(2018{\natexlab{a}}){Forgan}, {Hall}, {Meru}, \&
  {Rice}}]{Forgan_popsyn2}
{Forgan}, D.~H., {Hall}, C., {Meru}, F., \& {Rice}, W.~K.~M.
  2018{\natexlab{a}}, \mnras, 474, 5036

\bibitem[{{Forgan} {et~al.}(2016{\natexlab{b}}){Forgan}, {Ilee}, {Cyganowski},
  {Brogan}, \& {Hunter}}]{Forgan_massivestars}
{Forgan}, D.~H., {Ilee}, J.~D., {Cyganowski}, C.~J., {Brogan}, C.~L., \&
  {Hunter}, T.~R. 2016{\natexlab{b}}, \mnras, 463, 957

\bibitem[{{Forgan} {et~al.}(2018{\natexlab{b}}){Forgan}, {Ram{\'o}n-Fox}, \&
  {Bonnell}}]{tache}
{Forgan}, D.~H., {Ram{\'o}n-Fox}, F.~G., \& {Bonnell}, I.~A.
  2018{\natexlab{b}}, ArXiv e-prints, arXiv:1802.01364

\bibitem[{{Fukagawa} {et~al.}(2006){Fukagawa}, {Tamura}, {Itoh}, {Kudo},
  {Imaeda}, {Oasa}, {Hayashi}, \& {Hayashi}}]{Fukagawa_tidal_tail}
{Fukagawa}, M., {Tamura}, M., {Itoh}, Y., {et~al.} 2006, \apjl, 636, L153

\bibitem[{{Fung} \& {Dong}(2015)}]{Fung2015}
{Fung}, J., \& {Dong}, R. 2015, \apjl, 815, L21

\bibitem[{{Galvagni} \& {Mayer}(2014)}]{Galvagni_pop_syn}
{Galvagni}, M., \& {Mayer}, L. 2014, \mnras, 437, 2909

\bibitem[{{Gammie}(2001)}]{Gammie_betacool}
{Gammie}, C.~F. 2001, \apj, 553, 174

\bibitem[{{Goodman} \& {Rafikov}(2001)}]{goodman_2001}
{Goodman}, J., \& {Rafikov}, R.~R. 2001, \apj, 552, 793

\bibitem[{{Greaves} \& {Rice}(2010)}]{Greaves_Rice2010}
{Greaves}, J.~S., \& {Rice}, W.~K.~M. 2010, \mnras, 407, 1981

\bibitem[{{Hall} {et~al.}(2016){Hall}, {Forgan}, {Rice}, {Harries}, {Klaassen},
  \& {Biller}}]{hall_2016}
{Hall}, C., {Forgan}, D., {Rice}, K., {et~al.} 2016, \mnras, 458, 306

\bibitem[{{Hall} {et~al.}(2018){Hall}, {Rice}, {Dipierro}, {Forgan}, {Harries},
  \& {Alexander}}]{Hall2018}
{Hall}, C., {Rice}, K., {Dipierro}, G., {et~al.} 2018, \mnras, 477, 1004

\bibitem[{{Haworth} {et~al.}(2017){Haworth}, {Glover}, {Koepferl}, {Bisbas}, \&
  {Dale}}]{Haworth2017}
{Haworth}, T.~J., {Glover}, S.~C.~O., {Koepferl}, C.~M., {Bisbas}, T.~G., \&
  {Dale}, J.~E. 2017, ArXiv e-prints, arXiv:1711.05275

\bibitem[{{Isella} {et~al.}(2009){Isella}, {Carpenter}, \&
  {Sargent}}]{isella_2009_structure}
{Isella}, A., {Carpenter}, J.~M., \& {Sargent}, A.~I. 2009, \apj, 701, 260

\bibitem[{{Laughlin} \& {Bodenheimer}(1994)}]{Laughlin_Bodenheimer_t_rad}
{Laughlin}, G., \& {Bodenheimer}, P. 1994, \apj, 436, 335

\bibitem[{{Lin} \& {Shu}(1964)}]{LinShu1964}
{Lin}, C.~C., \& {Shu}, F.~H. 1964, \apj, 140, 646

\bibitem[{{Lin} \& {Papaloizou}(1986)}]{Lin_Papaloizou_TypeII}
{Lin}, D.~N.~C., \& {Papaloizou}, J. 1986, \apj, 309, 846

\bibitem[{{Lodato} \& {Rice}(2005)}]{Lodato_Rice_massive_disc}
{Lodato}, G., \& {Rice}, W.~K.~M. 2005, \mnras, 358, 1489

\bibitem[{{Luhman} \& {Rieke}(1999)}]{rho_oph_spectraltype_age}
{Luhman}, K.~L., \& {Rieke}, G.~H. 1999, \apj, 525, 440

\bibitem[{{Meru}(2015)}]{Triggered}
{Meru}, F. 2015, \mnras, 454, 2529

\bibitem[{{Meru} {et~al.}(2017){Meru}, {Juh{\'a}sz}, {Ilee}, {Clarke},
  {Rosotti}, \& {Booth}}]{meru_elias}
{Meru}, F., {Juh{\'a}sz}, A., {Ilee}, J.~D., {et~al.} 2017, \apjl, 839, L24

\bibitem[{{Muto} {et~al.}(2012{\natexlab{a}}){Muto}, {Grady}, {Hashimoto},
  {Fukagawa}, {Hornbeck}, {Sitko}, {Russell}, {Werren}, {Cur{\'e}}, {Currie},
  {Ohashi}, {Okamoto}, {Momose}, {Honda}, {Inutsuka}, {Takeuchi}, {Dong},
  {Abe}, {Brandner}, {Brandt}, {Carson}, {Egner}, {Feldt}, {Fukue}, {Goto},
  {Guyon}, {Hayano}, {Hayashi}, {Hayashi}, {Henning}, {Hodapp}, {Ishii}, {Iye},
  {Janson}, {Kandori}, {Knapp}, {Kudo}, {Kusakabe}, {Kuzuhara}, {Matsuo},
  {Mayama}, {McElwain}, {Miyama}, {Morino}, {Moro-Martin}, {Nishimura}, {Pyo},
  {Serabyn}, {Suto}, {Suzuki}, {Takami}, {Takato}, {Terada}, {Thalmann},
  {Tomono}, {Turner}, {Watanabe}, {Wisniewski}, {Yamada}, {Takami}, {Usuda}, \&
  {Tamura}}]{Muto_disc_structure}
{Muto}, T., {Grady}, C.~A., {Hashimoto}, J., {et~al.} 2012{\natexlab{a}},
  \apjl, 748, L22

\bibitem[{{Muto} {et~al.}(2012{\natexlab{b}}){Muto}, {Grady}, {Hashimoto},
  {Fukagawa}, {Hornbeck}, {Sitko}, {Russell}, {Werren}, {Cur{\'e}}, {Currie},
  {Ohashi}, {Okamoto}, {Momose}, {Honda}, {Inutsuka}, {Takeuchi}, {Dong},
  {Abe}, {Brandner}, {Brandt}, {Carson}, {Egner}, {Feldt}, {Fukue}, {Goto},
  {Guyon}, {Hayano}, {Hayashi}, {Hayashi}, {Henning}, {Hodapp}, {Ishii}, {Iye},
  {Janson}, {Kandori}, {Knapp}, {Kudo}, {Kusakabe}, {Kuzuhara}, {Matsuo},
  {Mayama}, {McElwain}, {Miyama}, {Morino}, {Moro-Martin}, {Nishimura}, {Pyo},
  {Serabyn}, {Suto}, {Suzuki}, {Takami}, {Takato}, {Terada}, {Thalmann},
  {Tomono}, {Turner}, {Watanabe}, {Wisniewski}, {Yamada}, {Takami}, {Usuda}, \&
  {Tamura}}]{muto_2012}
---. 2012{\natexlab{b}}, \apjl, 748, L22

\bibitem[{{Natta} {et~al.}(2006){Natta}, {Testi}, \&
  {Randich}}]{rho_oph_accretion}
{Natta}, A., {Testi}, L., \& {Randich}, S. 2006, \aap, 452, 245

\bibitem[{{P{\'e}rez} {et~al.}(2016){P{\'e}rez}, {Carpenter}, {Andrews},
  {Ricci}, {Isella}, {Linz}, {Sargent}, {Wilner}, {Henning}, {Deller},
  {Chandler}, {Dullemond}, {Lazio}, {Menten}, {Corder}, {Storm}, {Testi},
  {Tazzari}, {Kwon}, {Calvet}, {Greaves}, {Harris}, \&
  {Mundy}}]{Elias227_Science}
{P{\'e}rez}, L.~M., {Carpenter}, J.~M., {Andrews}, S.~M., {et~al.} 2016,
  Science, 353, 1519

\bibitem[{{Rafikov}(2002)}]{rafikov_2002}
{Rafikov}, R.~R. 2002, \apj, 569, 997

\bibitem[{{Ricci} {et~al.}(2010){Ricci}, {Testi}, {Natta}, \&
  {Brooks}}]{Ricci_mm_cm_rhoOph}
{Ricci}, L., {Testi}, L., {Natta}, A., \& {Brooks}, K.~J. 2010, \aap, 521, A66

\bibitem[{{Rice} {et~al.}(2005){Rice}, {Lodato}, \&
  {Armitage}}]{Rice_beta_condition}
{Rice}, W.~K.~M., {Lodato}, G., \& {Armitage}, P.~J. 2005, \mnras, 364, L56

\bibitem[{{Stamatellos} {et~al.}(2007){Stamatellos}, {Hubber}, \&
  {Whitworth}}]{Stamatellos_BD_formation}
{Stamatellos}, D., {Hubber}, D.~A., \& {Whitworth}, A.~P. 2007, \mnras, 382,
  L30

\bibitem[{{Tomida} {et~al.}(2017){Tomida}, {Machida}, {Hosokawa}, {Sakurai}, \&
  {Lin}}]{Tomida_Elias227_GI}
{Tomida}, K., {Machida}, M.~N., {Hosokawa}, T., {Sakurai}, Y., \& {Lin}, C.~H.
  2017, \apjl, 835, L11

\bibitem[{{Toomre}(1964)}]{Toomre_stability1964}
{Toomre}, A. 1964, \apj, 139, 1217

\bibitem[{{Whitehouse} \& {Bate}(2006)}]{WH_Bate_science}
{Whitehouse}, S.~C., \& {Bate}, M.~R. 2006, \mnras, 367, 32

\bibitem[{{Whitehouse} {et~al.}(2005){Whitehouse}, {Bate}, \&
  {Monaghan}}]{WH_Bate_Monaghan2005}
{Whitehouse}, S.~C., {Bate}, M.~R., \& {Monaghan}, J.~J. 2005, \mnras, 364,
  1367

\bibitem[{{Zhu} {et~al.}(2015){Zhu}, {Dong}, {Stone}, \& {Rafikov}}]{zhu_2015}
{Zhu}, Z., {Dong}, R., {Stone}, J.~M., \& {Rafikov}, R.~R. 2015, \apj, 813, 88

\end{thebibliography}

%% This command is needed to show the entire author+affilation list when
%% the collaboration and author truncation commands are used.  It has to
%% go at the end of the manuscript.
%\allauthors

%% Include this line if you are using the \added, \replaced, \deleted
%% commands to see a summary list of all changes at the end of the article.
%\listofchanges

\end{document}